\documentclass[prl, twocolumn,preprintnumbers,amsmath,amssymb ]{revtex4}

\usepackage{graphicx}
\usepackage{dcolumn}
\usepackage{bm}

\begin{document}

\input epsf.sty



\title{Magnetic vortices instead of stripes: another interpretation of magnetic neutron scattering in lanthanum cuprates}

\author{Boris V. Fine}

\affiliation{ Department of Physics and Astronomy, 101 South College, 1413 Circle Drive, Knoxville, Tennessee 37996, USA}

\date{October 26, 2006}

\begin{abstract}
It is proposed that a two-dimensional magnetic superstructure closely related to the one mentioned recently by Christensen {\it et al.} constitutes a viable interpretation of the fourfold splitting of the magnetic $(\pi, \pi)$ peak in lanthanum cuprates. (This splitting is usually interpreted as evidence for stripes.) The superstructure in question has the topology of a square crystal of magnetic vortices with approximate periodicity $4a \times 4a$. This vortex crystal exhibits no magnetic antiphase lines. It is shown that such a superstructure is magnetically stable in the approximation of staggered spin polarizations and that it should be accompanied by charge modulation characterized by charge peaks at the positions observed experimentally.
\end{abstract}


\maketitle

\narrowtext
\pagebreak

This communication is motivated by the recent work of Christensen {\it et al.}\cite{Christensen-etal-07}  reporting significant new results of magnetic neutron scattering in La$_{1.48}$Nd$_{0.4}$Sr$_{0.12}$CuO$_4$. The above authors interpret their results as establishing ``beyond reasonable doubt'' that the magnetic order in La$_{1.48}$Nd$_{0.4}$Sr$_{0.12}$CuO$_4$ is one-dimensionally (1D) modulated. They nevertheless give an example of a noncollinear two-dimensional (2D) modulation, which they state would be consistent with experiment, but which they rule out as unphysical. However, the reason for such a conclusion seems to be that the above 2D modulation contains unnecessary elements obscuring the basic idea of the authors to consider the coherent superposition of two 1D modulations with orthogonal wave vectors {\it and  orthogonal spin polarizations}. As shown in Fig.1, a straightforward implementation of this idea reveals an interesting magnetic topology, which has not been discussed so far in the context of cuprates: namely, the square lattice of magnetic vortices. The relevant picture in Ref.\cite{Christensen-etal-07}  contains lines of nonmagnetic sites connecting the vortex cores. These lines mask the vortex nature of the pattern and make it look unphysical. Figure 1, however, highlights the simple fact that a 2D interpretation of the fourfold splitting of the magnetic neutron peak at $(\pi, \pi)$\cite{Tranquada-etal-95, Fujita-etal-04} does not necessarily lead to antiphase magnetic lines (the lines of zero or nearly zero spin polarization, along which the antiferromagnetic (AF) order changes sign). The antiphase lines are unavoidable, only when one limits oneself to a collinear 2D superstructure\cite{Tranquada-etal-99,Fine-hitc-prb04,Fine-NQR-prb07} referred to as a grid, or checkerboard [see Fig.~2(a)]. The result of Christensen et al., indeed,  constitutes a strong piece of evidence against a grid, but not against a magnetic vortex lattice. The purpose of the present Rapid Communication is to clarify this matter and to discuss the most obvious properties of the latter superstructure.


\begin{figure} \setlength{\unitlength}{0.1cm}
\begin{picture}(100, 181) 
{ 
\put(2, 176){{\Large (a)}}
\put(0, 134){ \epsfxsize= 1.4in \epsfbox{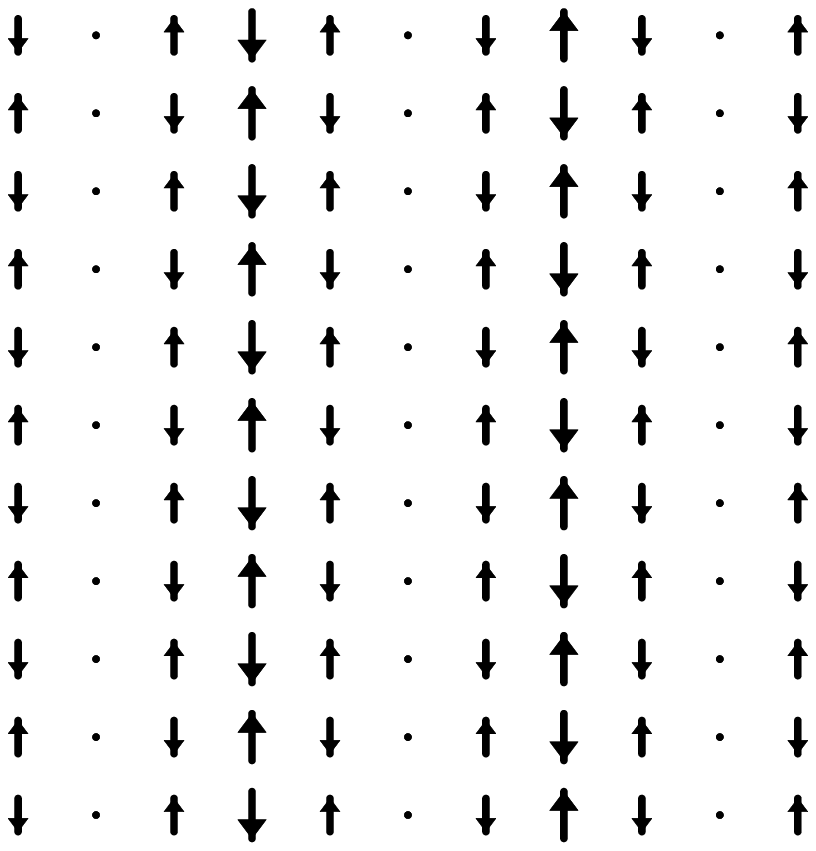  } }
\put(42, 151){{\Large +}}
\put(48, 136){ \epsfxsize= 1.4in \epsfbox{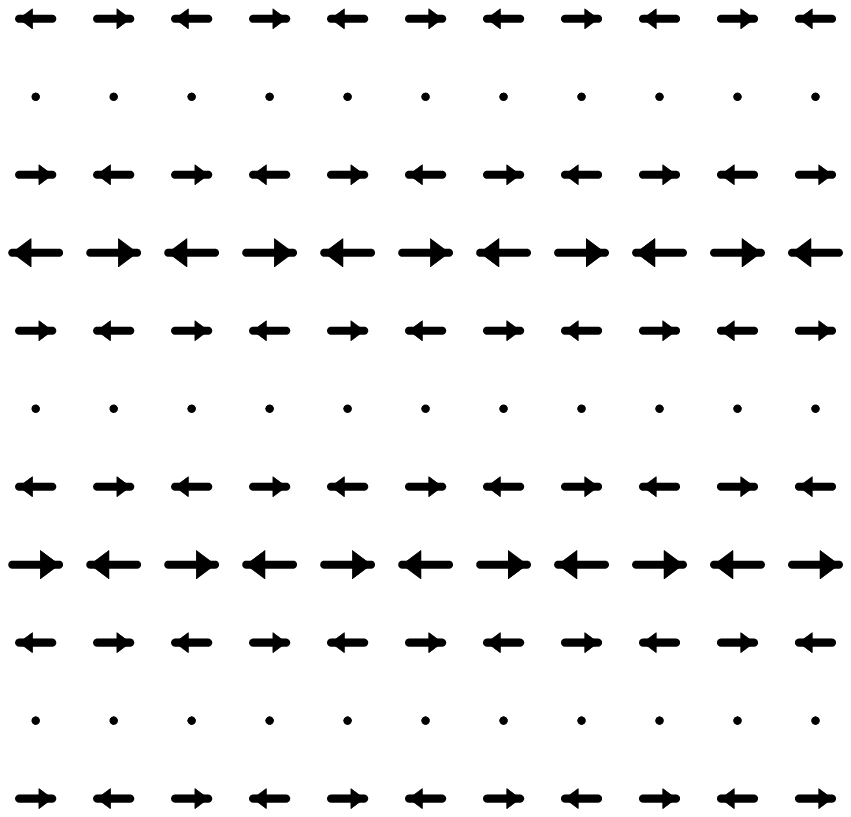  } }
\put(5, 93){{\Large =}}
\put(13, 67){ \epsfxsize= 2.3in \epsfbox{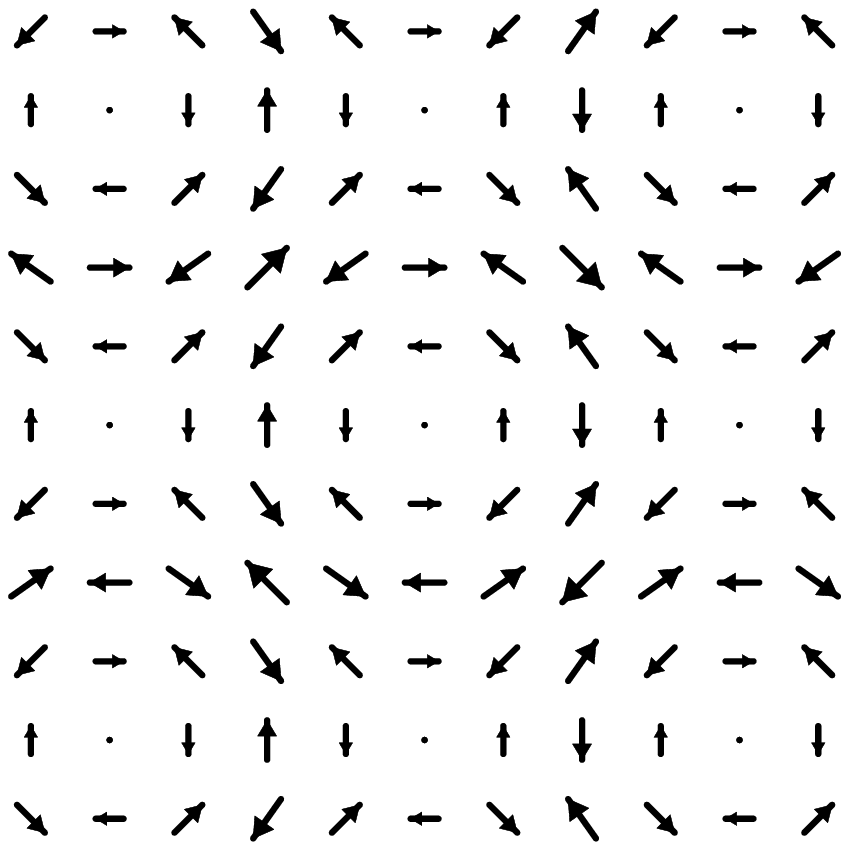  } }
\put(2, 51){{\Large (b)}}
\put(13, 0){ \epsfxsize= 2.3in \epsfbox{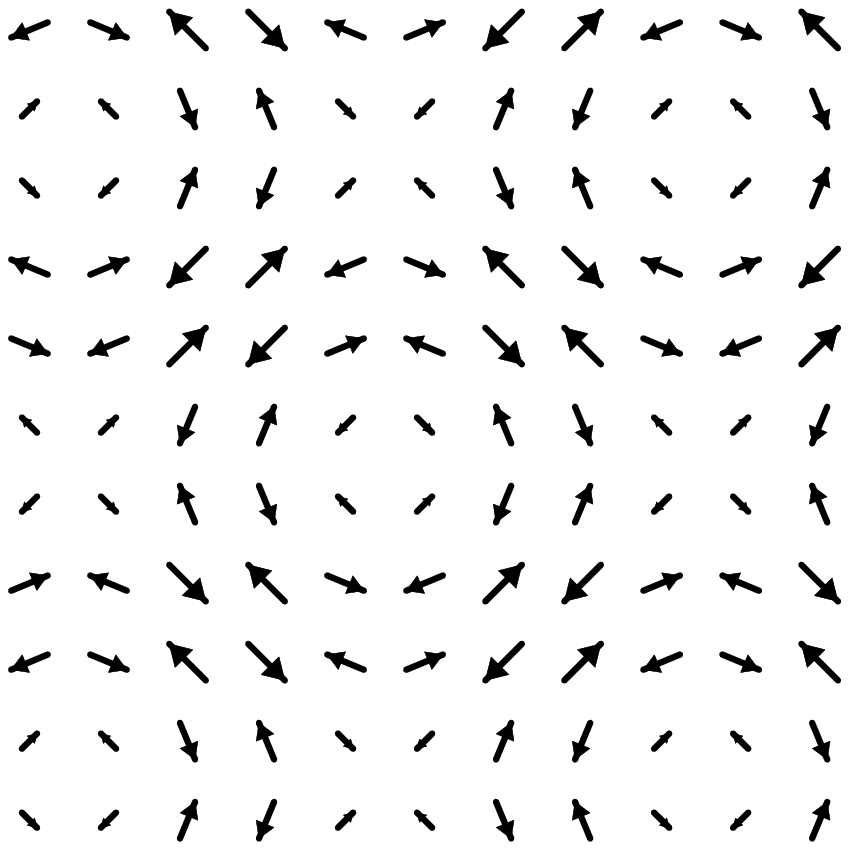  } }
}
\end{picture} 
\caption{(a) Magnetic (site-centered) vortex lattice produced by the coherent superposition of two one-dimensional spin harmonics (shown above). (b) Bond-centered version of the same vortex lattice.
} 
\label{fig-main} 
\end{figure}


Figure 1 contains two versions of the magnetic vortex lattices having spin polarizations consistent with the experiment of Ref.\cite{Christensen-etal-07}. These lattices are commensurate and can be labeled as ``site-centered'' [Fig.~1(a)], and ``bond-centered'' [Fig.~1(b)]. The experiment indicates that the magnetic modulation is incommensurate, but the following discussion will be amenable to this case.
For, completeness, Fig.~2 represents three other modulations: grid, ``radial'' vortex lattice, and the coherent superposition of two helical harmonics---all of which appear to be inconsistent with the experiment of Ref.\cite{Christensen-etal-07}, but can, possibly, exist as fluctuations at higher energies. In particular, it is worth noting that the  superposition of two helical harmonics shown in Fig.~2(c) should be accompanied by 1D diagonal charge modulation.

Now the question arises whether the superstructures shown in Fig.1 are unphysical. This is not obvious at all. Breaking the 2D AF order via the creation of vortex pairs is the well-known mechanism of the thermal Berezinskii-Kosterlitz-Thouless transition\cite{Berezinskii-71,Kosterlitz-etal-73}. It is, therefore, conceivable that doping can induce similar physics. Unlike vortices induced by temperature,  the doping-induced magnetic vortices should have cores, which attract doped charge carriers. This, in turn, leads to the repulsion between vortices. As a result, it is easy to imagine that mobile magnetic vortices exhibiting  Wigner-crystal-like correlations exist generically in cuprates, but only in 1/8-doped lanthanum cuprates do these vortices freeze into an actual crystal. Some very relevant theoretical studies of magnetic vortices with charged cores occupied by single holes were reported in Refs.\cite{Seibold-98,Berciu-etal-99,Timm-etal-99,Timm-etal-00}. However, if the spin textures shown in Fig.~1 are actually present in 1/8-doped cuprates, they contain two holes per magnetic vortex.

It is possible to show that the vortex crystal shown in Fig.~1 is  stable magnetically in the case of nearest-neighbor Heisenberg exchange in the approximation of staggered spin polarizations; i.e. an exotic interaction is not necessary to stabilize it.
The energy of the nearest-neighbor Heisenberg interaction on a square lattice is:
\begin{equation}
E = J \sum^{_{\hbox{\small NN}}}_{ij,m>i,n>j} \mathbf S_{ij} \cdot \mathbf S_{mn}
\label{E}
\end{equation}
where $(ij)$ and $(mn)$ are the pairs of square lattice indices, $\mathbf S_{ij}$ are the local staggered spin polarizations, and $J$ is the [positive] exchange constant. Superscript ``NN'' implies that sites $(mn)$ are the nearest neighbors of sites $(ij)$. According to Eq.(\ref{E}) each spin should experience a local field:
\begin{equation}
\mathbf h_{ij} = J \sum^{_{\hbox{\small NN}}}_{mn} \mathbf S_{mn}.
\label{h}
\end{equation}
The locally stable configuration should have
$\mathbf h_{ij} \downarrow \uparrow \mathbf S_{ij}$ for each lattice site, which can be shown as follows.The relevant (nonhelical) spin harmonics can be presented as
\begin{equation}
\mathbf S_{ij} = (-1)^{i+j}
\left[
\mathbf S_{q_1} \hbox{sin} \left( \mathbf q_1 \cdot \mathbf r_{ij} \right)
+
\mathbf S_{q_2} \hbox{sin} \left( \mathbf q_2 \cdot \mathbf r_{ij} \right)
\right],
\label{S}
\end{equation}
where $\mathbf q_1 = (q_0, 0)$ and $\mathbf q_2 = (0, q_0)$ are the wave vectors of the two harmonics and $\mathbf S_{q_1}$ and $\mathbf S_{q_2}$ are the corresponding polarization amplitudes, $\mathbf r_{ij}$ are the radius vectors of the lattice sites,  $q_0 \approx {\pi \over 4 a}$, and $a$ is the lattice period. 
The experiment of Ref.\cite{Christensen-etal-07}indicates that $\mathbf S_{q_1} \bot \mathbf q_1$ and $\mathbf S_{q_2} \bot \mathbf q_2$.
The substitution of Eq.(\ref{S}) into Eq.(\ref{h}) gives 
\begin{equation}
\mathbf h_{ij} = - 4 \ J \ \hbox{cos}^2 \left({q_0 a \over 2} \right) \  \mathbf S_{ij},
\label{hS}
\end{equation}
which is, indeed antiparallel to $\mathbf S_{ij}$ for each lattice site. 

Equation(\ref{hS}) is, in fact, valid for an arbitrary relative orientation of $\mathbf S_{q_1}$ and $\mathbf S_{q_2}$ with respect to each other and with respect to $\mathbf q_1$ and $\mathbf q_2$ and also for the superposition of two helical modes with wave vectors $\mathbf q_1$ and $\mathbf q_2$.
This means that all superstructures shown in Figs.~1 and 2 are locally stable. Yet the vortex lattice appears to be more stable magnetically than the diagonal grid. A diagonal grid divides the system into AF clusters, which can rotate as a whole without increasing (or with decreasing) the magnetic energy of the system. The vortex lattice does not afford such cluster rotations.


\begin{figure} \setlength{\unitlength}{0.1cm}
\begin{picture}(100, 196) 
{ 
\put(6, 190){{\Large (a)}}
\put(13, 134){ \epsfxsize= 2.3in \epsfbox{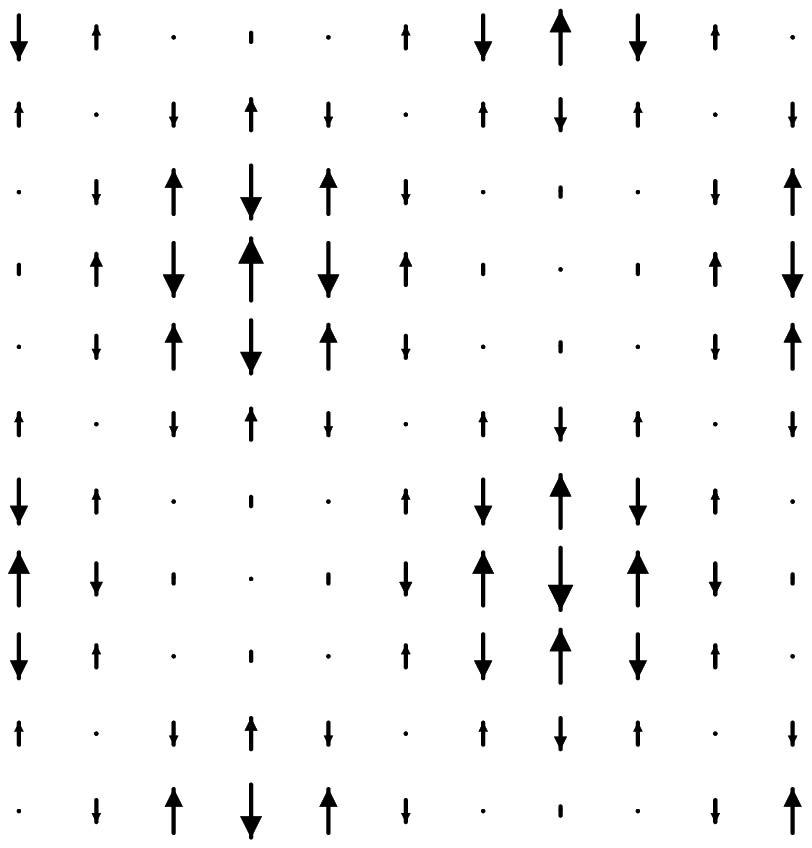  } }
\put(6, 119){{\Large (b)}}
\put(13, 67){ \epsfxsize= 2.3in \epsfbox{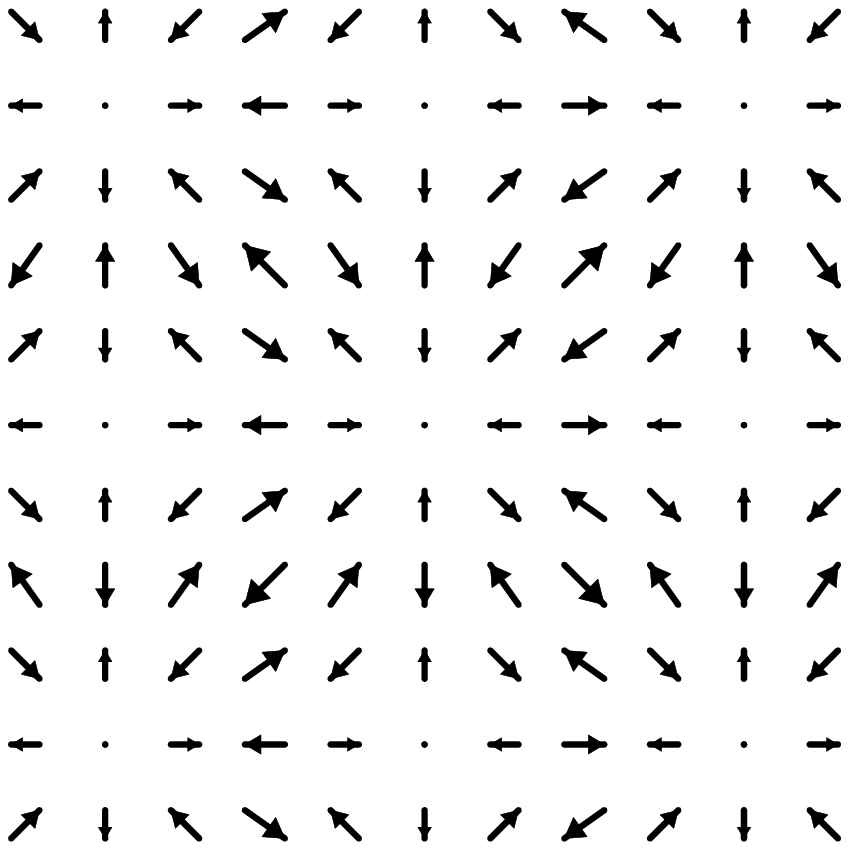  } }
\put(6, 51){{\Large (c)}}
\put(13, 0){ \epsfxsize= 2.3in \epsfbox{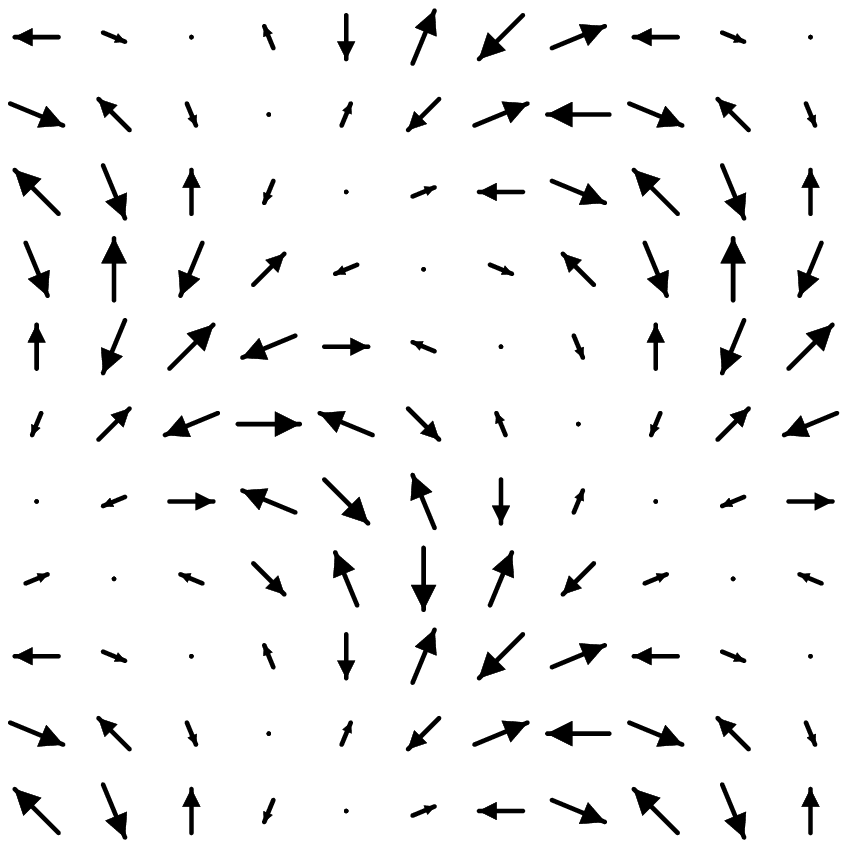  } }
}
\end{picture} 
\caption{(a) Diagonal grid superstructure. (b) ``Radial'' vortex lattice (obtained from Fig.~1(a) by rotating all spin polarizations by 90 degrees). (c) Coherent superposition of two one-dimensional helical harmonics.
} 
\label{fig-other} 
\end{figure}


Using Eqs.(\ref{E}) and (\ref{S}) it is also possible to express the total magnetic energy of the system as
\begin{equation}
E = - \ J \ N \  \hbox{cos}^2 \left({q_0 a \over 2} \right) \ \ 
\left(\mathbf S_{q_1}^2 + \mathbf S_{q_2}^2 \right),
\label{E1}
\end{equation}
where $N$ is the total number of lattice sites. Like Eq.(\ref{hS}), Eq. (\ref{E1}) holds for any mutual orientation of $\mathbf S_{q_1}$ and $\mathbf S_{q_2}$. Therefore, by changing the relative angle of $\mathbf S_{q_1}$ and $\mathbf S_{q_2}$, it is possible to produce a whole family of locally stable 2D spin modulations having the same exchange energy.  This family ranges from vortex lattices ($\mathbf S_{q_1} \bot  \mathbf S_{q_2}$) to diagonal grid ($\mathbf S_{q_1} ||  \mathbf S_{q_2}$). Such an unusual degeneracy should presumably be lifted, once the energies of doped charge carriers are taken into account.

It is quite obvious from Fig.~1 that the magnetic vortex lattice should be accompanied by an approximate $4a \times 4a$ modulation of the charge density similar to the one observed by scanning tunneling spectroscopy in other cuprate families\cite{Hoffman-etal-02,Vershinin-etal-04,Hanaguri-etal-04}. Technically, this modulation can be obtained as follows: In the Landau-type expansion\cite{Zachar-etal-98}, the local charge density $\rho_{ij}$ should couple to the square of the local spin polarization $\mathbf S_{ij}$ and, thereby, become proportional to it, i.e.
\begin{eqnarray}
\rho_{ij} \cong \mathbf S_{ij}^2 &=&
\mathbf S_{q_1}^2 \  \hbox{sin}^2 \left( \mathbf q_1 \cdot \mathbf r_{ij} \right)
\nonumber
\\
&&
+ \ 
2 \ (\mathbf S_{q_1} \cdot \mathbf S_{q_2}) \ 
\hbox{sin} \left( \mathbf q_1 \cdot \mathbf r_{ij} \right) \ 
\hbox{sin} \left( \mathbf q_2 \cdot \mathbf r_{ij} \right)
\nonumber
\\
&&
+ \ 
\mathbf S_{q_2}^2 \  \hbox{sin}^2 \left( \mathbf q_2 \cdot \mathbf r_{ij} \right).
\label{rho}
\end{eqnarray}
In the case of the grid superstructure (i.e. with
$\mathbf S_{q_1} || \mathbf S_{q_2}$), the above modulation has the leading harmonic corresponding to the second term on the right hand side. Therefore, in this approximation, one expects leading charge modulation peaks to be rotated by 45 degrees with respect to the spin modulation peaks\cite{Tranquada-etal-99,Kivelson-etal-03,Robertson-etal-06}. However, in the case of the vortex lattice, $\mathbf S_{q_1} \bot  \mathbf S_{q_2}$, and, as a result,  the second term vanishes, leaving the first and third terms corresponding to wave vectors $(\pm 2 q_0, 0)$ and $(0, \pm 2 q_0)$  as leading harmonics. These harmonics are not rotated with respect to the magnetic ones and, in fact, have the same wave vectors as those of the charge peaks observed experimentally\cite{Tranquada-etal-95,Zimmermann-etal-98,Tranquada-etal-99,Fujita-etal-04}. Landau expansion then predicts no other charge harmonics for the vortex lattice.

Beyond the Landau expansion, one should be concerned with the amplitude of charge modulation and the resulting cost in terms of the Coulomb energy. This author has previously argued\cite{Fine-hitc-prb04,Fine-NQR-prb07} that, in the case of a diagonal grid, the Coulomb repulsion can conceivably lead to the modification of the Landau expansion results by pushing the charge density away from the intersections of magnetic antiphase lines. In that case, there are four doped holes per intersection and also the local magnetic energy cost of moving a hole along an antiphase line is zero. In the case of the vortex lattice with approximate periodicity $4a \times 4a$, there are two doped holes per vortex (around doping level 1/8); i.e. less Coulomb energy is involved, and, besides, moving a hole away from a vortex core in any direction has local magnetic energy cost. Therefore, it is less likely than in the case of the grid (but not impossible), that Coulomb repulsion would lead to a shift of the leading charge peaks. Coulomb repulsion or other higher order terms can also lead to new charge peaks, which, if observed away from the principal lattice directions, would falsify the 1D stripe interpretation\cite{Kivelson-etal-03,Robertson-etal-06}.

Like the diagonal grid, the magnetic vortex lattice does not have a preferred orientation to couple to the crystal anisotropy of the low temperature tetragonal (LTT) phase exhibited by most materials, where the static spin modulations were observed. However, the role of the LTT phase can be the following: The transition to the LTT phase takes place from the nearby low temperature orthorhombic (LTO) phase as a function of temperature and doping. This transition is of the first order, because neither of these two phases can be obtained from the other by spontaneous symmetry breaking\cite{Fine-Egami-07}. The first-order phase transition as a function of charge carrier concentration (doping) cannot proceed homogeneously, because, in the vicinity of the homogeneous phase transition, the system becomes unstable towards phase separation.  Since the transition is of the first order, the phase separation cannot be suppressed completely by the Coulomb energy associated with the uncompensated charge, but the Coulomb factor will certainly limit the scale and charge amplitude of the resulting inhomogeneous pattern\cite{Emery-etal-93}. Since the energy difference between the LTO and LTT phases should not be large, the LTO-LTT transition alone is unlikely to induce a significant amplitude of charge modulations. However, if the system is predisposed to a nanoscale phase separation, as seems to be the case, then the proximity to the LTO-LTT transition can help to stabilize both the one-dimensional and two-dimensional charge modulations.

Two holes moving inside a magnetic vortex would be consistent at some level with various superconductivity models involving preformed pairs alone or coupled to a second fermionic component. In particular, the two-component superconductivity model proposed earlier by this author\cite{Fine-hitc-prb04,Fine-delta-prl05}  in the context of a grid hypothesis should be amenable to the case of magnetic vortex lattice by changing the topology of the second fermionic component.

To conclude, the magnetic vortex lattice appears to be a viable interpretation of the fourfold splitting of the neutron $(\pi, \pi)$ peak in lanthanum cuprates.

The author is grateful to T. Egami for discussions related to this work.


\end{document}